\documentclass[11pt,twoside]{article}
\pdfoutput=1

\usepackage{asp2006}
\usepackage{graphicx}

\begin{document}
\title{On the Role of Acoustic-gravity Waves in the Energetics of the Solar Atmosphere} 
\author{Thomas Straus\altaffilmark{1}, Bernhard Fleck\altaffilmark{2}, Stuart M. Jefferies\altaffilmark{3}, 
Scott W. McIntosh\altaffilmark{4}, 
Giuseppe Severino\altaffilmark{1}, Matthias Steffen\altaffilmark{5}, Theodore D. Tarbell\altaffilmark{6}}   
\altaffiltext{1}{INAF / Osservatorio Astronomico di Capodimonte, Via Moiariello 16, 80131 Napoli, Italy}
\altaffiltext{2}{ESA Science Operations Dep., c/o NASA GSFC, Mailcode 671.1, Greenbelt, MD 20771, USA}
\altaffiltext{3}{Institute for Astronomy, University of Hawaii, 34 Ohia Ku Street, Pukalani, HI 96768, USA}
\altaffiltext{4}{High Altitude Observatory, NCAR, P.O. Box 3000, Boulder, CO 80307, USA}
\altaffiltext{5}{Astrophysikalisches Institut Potsdam, An der Sternwarte 16, D-14482 Potsdam, Germany}
\altaffiltext{6}{Lockheed Martin Solar and Astrophysics Laboratory, 3251 Hanover Street
Palo Alto, CA 94304}

\begin{abstract} 
In a recent paper \citep{SFJ} we determined the energy flux of internal gravity waves in the lower solar atmosphere using a combination of 3D numerical simulations and observations obtained with the IBIS instrument operated at the Dunn Solar Telescope and the Michelson Doppler Imager (MDI) on SOHO. In this paper we extend these studies using coordinated observations from SOT/NFI and SOT/SP on Hinode and MDI. The new measurements confirm that gravity waves are the dominant phenomenon in the quiet middle/upper photosphere and that they transport more mechanical energy than the high-frequency ($> 5$\,mHz) acoustic waves, even though we find an acoustic flux 3-5 times larger than the upper limit estimate of \citet{FC}. It therefore appears justified to reconsider the significance of (non-M)HD waves for the energy balance of the solar chromosphere.
\end{abstract}

Even sixty years after the two pioneering papers by \citet{Schwarzschild} and \citet{Biermann}, who independently proposed high frequency acoustic waves as a key agent, the origin and mechanism of the heating of stellar atmospheres and coronae remains one of the great unsolved problems in solar and stellar physics. Recent studies by \citeauthor{FC} (\citeyear{FC}, \citeyear{Fossum}), hereafter F\&C, and \citet{Carlsson} conclude that high frequency waves are not sufficient to heat the solar chromosphere. Others \citep[e.g.][]{Cuntz, Wedemeyer,Kalkofen} question these results and argue for high-frequency waves to play an important role.

Stably stratified atmospheres can support and propagate not only acoustic waves but also internal gravity waves. These have been proposed as an agent for the mechanical heating of stellar atmospheres more than 25 years ago \citep{Mihalas}. However, the difficulties associated with directly observing them in our closest star, the Sun, have resulted in their neglect in this matter. Recently, by combining high quality 2D observations and 3D numerical simulations, \citet{SFJ}, hereafter SFJ, were able to determine the height dependence of the energy flux of internal gravity waves in the lower solar atmosphere and found that at the base of the Sun's chromosphere it is around 5\,kW\,m$^{-2}$, which is comparable to the radiative losses of the entire chromosphere. 

Here we revisit the role of both internal gravity waves as well as high frequency acoustic waves in the dynamics and energetics of the Sun's atmosphere using high quality Doppler velocity measurements obtained with SOT/NFI and SOT/SP on Hinode and with MDI on SOHO. 

The study of wave propagation characteristics requires simultaneous measurements in at least two different heights in the solar atmosphere. Gravity waves have long periods and small spatial scales, and a proper horizontal wavenumber separation is essential for their study. They therefore require both high spatial resolution and a large field-of-view, which limits the achievable cycle time of current instrumentation. For high frequency waves, on the other hand, a short cycle time is the key requirement. We therefore developed different observing programs, one targeted at gravity waves, the other at high frequency waves.

To study gravity waves, we combined SOT/NFI filtergrams in Mg b$_2$ with simultaneous MDI high-resolution observations in the Ni line at 6768\,\AA. The data were obtained on 2007 October 20. Both time series extend over 12\,hours with a cadence of 40\,seconds and 60\,seconds and a spatial resolution of $0\farcs32$/pixel and $0\farcs61$/pixel for SOT/NFI and MDI, respectively. The NFI filtergrams were interpolated to the times of the MDI observations using cubic splines. Dopplergrams were determined from differencing the blue and red wing filtergrams. They were calibrated by comparing the cumulative probability distributions of the Dopplergrams made with the normal filter settings, with those made with settings shifted by 20\,m\AA\ to the blue. The resulting Dopplergrams were coregistered and mapped onto the lower resolution MDI data. Both data cubes $v(x,y,t)$ were subjected to a 3D Fourier transform yielding complex Fourier coefficients $\tilde{v}(k_x,k_y,\nu)$. Fig.~1 shows the phase difference spectrum $\Phi(k_h,\nu)$ between the Mg b$_2$ and Ni 6768 Doppler signals after azimuthal integration of the complex crosspower $\tilde{v}_{\mathrm{Mg}}(k_x,k_y,\nu)\,\tilde{v}^{*}_{\mathrm{Ni}}(k_x,k_y,\nu)$  over $k_h^2=k_x^2+k_y^2$. It beautifully confirms the presence of internal gravity waves in the upper photosphere (transition to green-yellow-red in the gravity wave regime), and, thanks to the high wavenumber and frequency resolution, reveals interesting ridge/interridge structures in the region of $p$-modes, which will be discussed in another paper. For a height difference of 600\,km between the two signal forming layers (derived from the slope of the linear phase delay in the high-frequency part of the phase spectrum), the phase spectrum yields the vertical phase velocity $v_{\mathrm{ph,z}}(k_h,\nu)$, which in turn yields the vertical group velocity $v_{\mathrm{g,z}} = - v_{\mathrm{ph,z}} \sin^2(\theta)$ ($\theta$ is the angle between the oscillation in the wave and the vertical, for details see SFJ). With the Mg b$_2$ power spectrum and the VAL density of $5.97\cdot 10^{-7}\,$kg\,m$^{-3}$ at 720\,km, the vertical energy flux carried by gravity waves at that height can be estimated as $\rho_0 \langle v^2\rangle v_{\mathrm{g,z}}(k_h,\nu)\approx$\,2\,kW\,m$^{-2}$, which is consistent with the results of SFJ (Fig. 2). 

\begin{figure}
%\plotfiddle{straus_smallfig1}{2.5cm}{0}{35}{35}{14}{-10}
\includegraphics[width=0.6\textwidth]{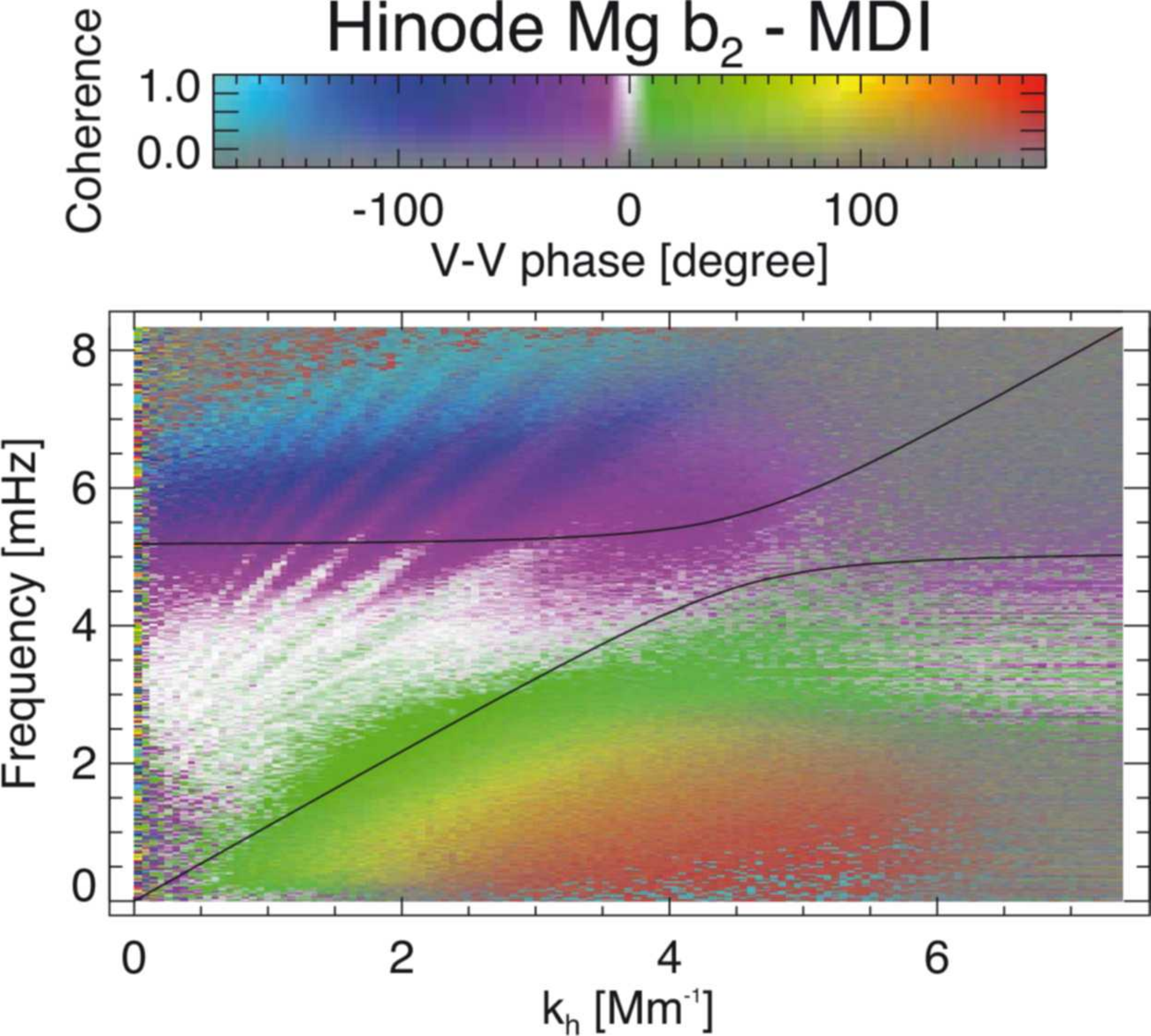}
\hfill\parbox[b]{0.45\textwidth}{
\caption{V-V phase difference spectrum $\Phi(k_h,\nu)$ between Mg b$_2$ and Ni\,{\sc i} 6768\,\AA. Note the area of downward phase propagation (green-yellow-red) in the region of propagating atmospheric gravity waves in the $k-\omega$-diagram.}}
\end{figure}

\begin{figure}
%\plotfiddle{straus_smallfig2}{5cm}{0}{45}{45}{-180}{-10}
\includegraphics[height=5cm]{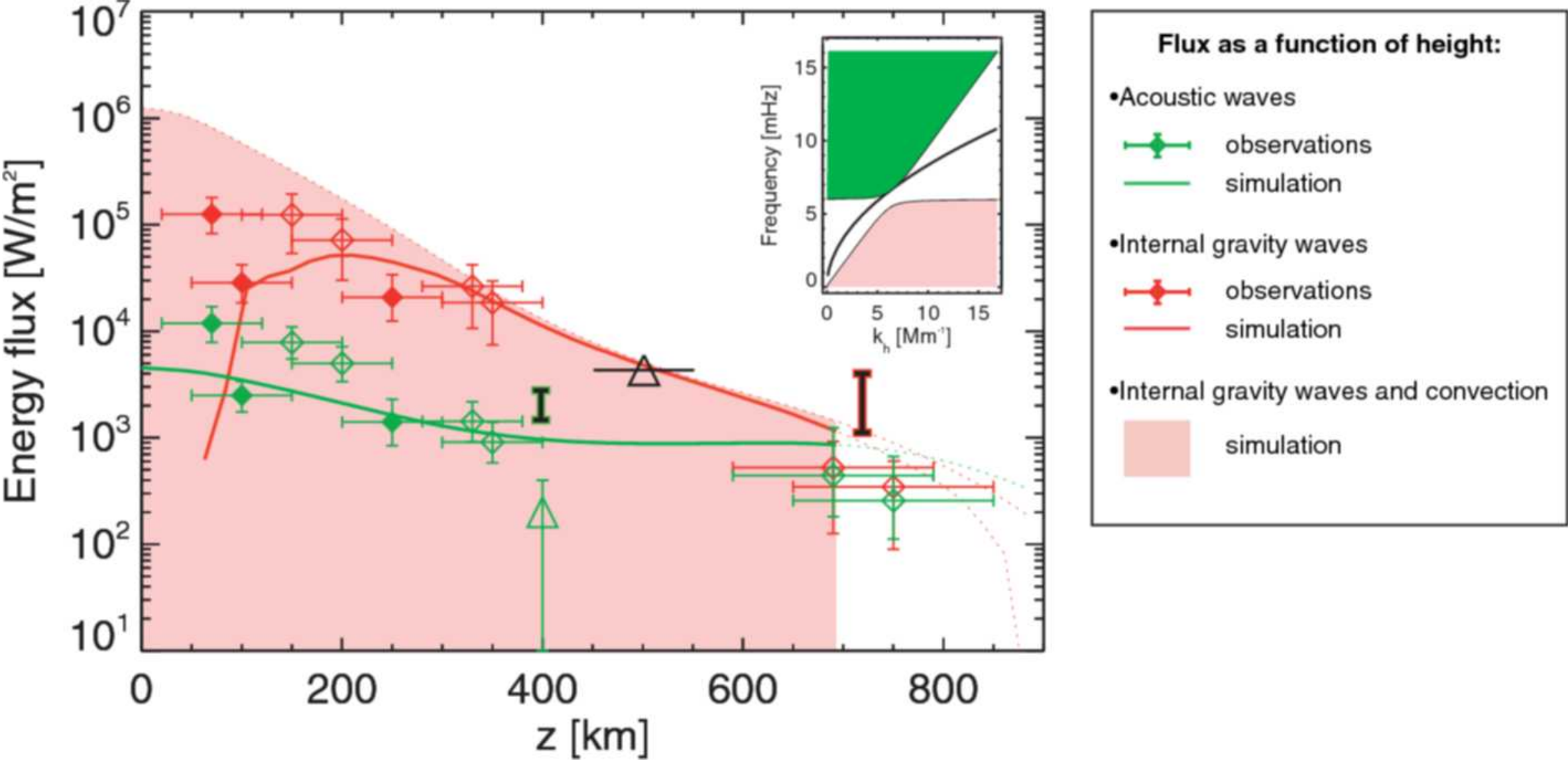}
\caption{Flux of atmospheric gravity waves (red) and high-frequency acoustic waves (green) in the solar atmosphere, as found by \protect\citet{SFJ} in numerical simulations (lines) and observations (symbols). 
The flux has been integrated over the regions in the $k_h-\nu$ diagram shown in the insert.
The new results obtained with Hinode in this work are added in bold. 
}
\end{figure}

\begin{figure}
%\plotfiddle{straus_smallfig3}{2.5cm}{0}{40}{40}{13}{0}
\includegraphics[width=0.6\textwidth]{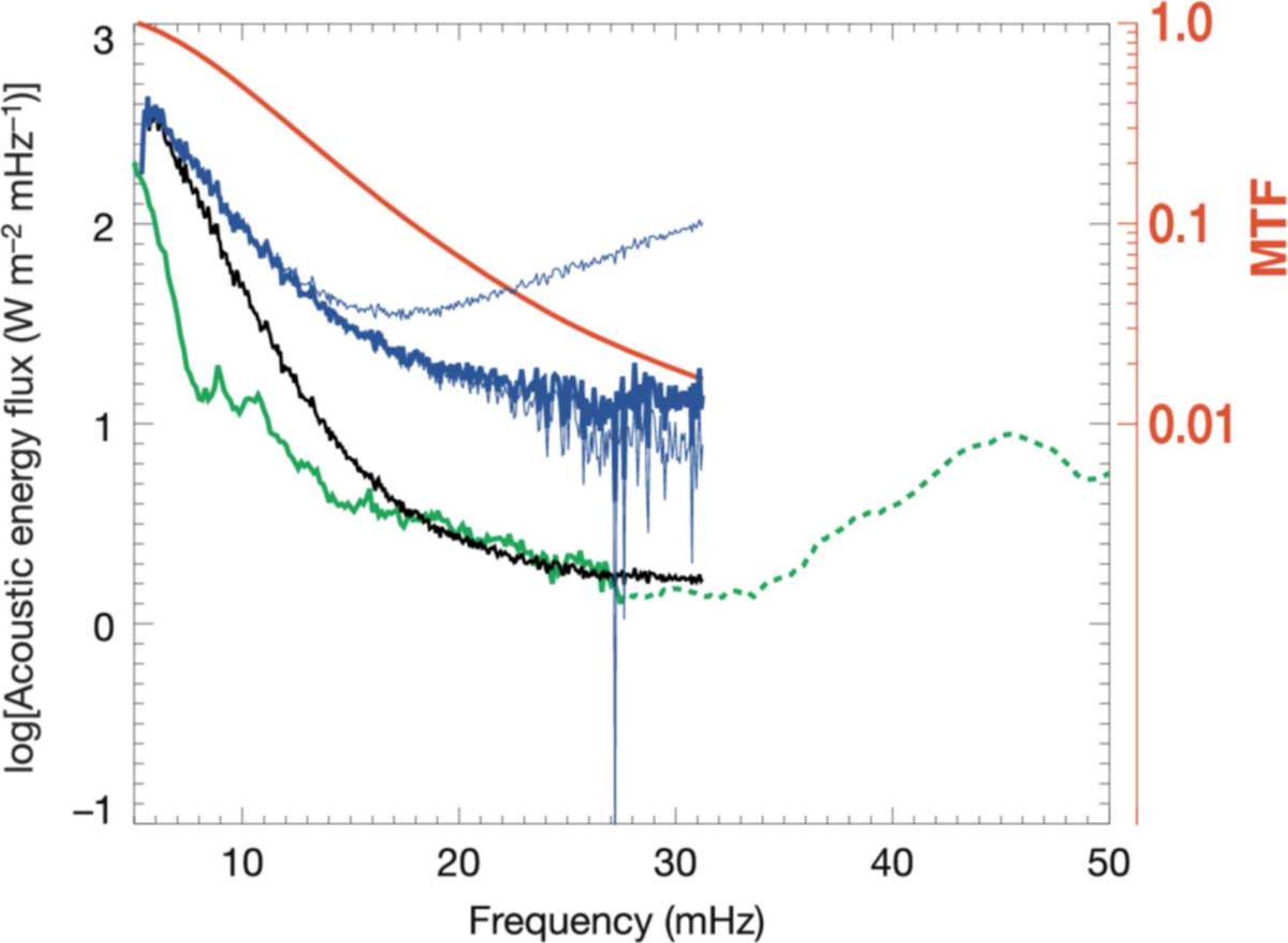}
\hfill\parbox[b]{0.46\textwidth}{
\caption{Flux of  high-frequency acoustic waves from observations with Hinode in the Fe 6301 line, before (black) and after (blue) correction for the line formation MTF (red). The flux given in F\&C is shown in green for comparison; see text.}}
\end{figure}

For the study of high frequency waves, we used two data sets obtained simultaneously with the spectropolarimeter (SP) and narrowband filter (NFI). For this short paper, we restrict ourselves to the discussion of the results from a data set obtained on 2007/08/13, and here on the results from the Fe\,{\sc i} 6301\,\AA\ line, the stronger one of the two Fe lines recorded with SP. The data set extends over 237 minutes at a cadence of 16 seconds. There are 1024 pixels along the slit with a pixel size of $0\farcs16$.
The Doppler signal of this line is formed at around 400\,km, i.e. close to the UV continuum analyzed by F\&C. A critically important factor in determining the energy flux of high frequency waves is the line formation modulation transfer function (MTF; \citeauthor{Keil}, \citeyear{Keil}). To better estimate this effect, we model a spectrum of monochromatic waves in a stratified atmosphere with frequencies in the range 6 to 100\,mHz and constant amplitude. With the synthesis of the Fe 6301 line profile the MTF can be determined as the ratio of the velocity power at the formation height of the line and the power of the Dopplershift velocity determined in the line core. The results are shown in Fig.~3, with the MTF in red, the measurements of the acoustic flux from F\&C in green, the ÒuncorrectedÓ flux determined from the power spectrum in black, and three MTF-corrected flux curves (assuming different S/N levels in the power tail) in blue. Note the steep fall-off of the MTF of this typical photospheric line by two orders of magnitude in the range $5<\nu<30\,$mHz. Depending on the noise level, taking into account MTF effects could even lead to a turnaround of power spectra at high frequencies. Using the thicker, middle blue curve (assumed S/N = 0.17), we estimate the flux of waves with frequencies between 5 and 31\,mHz at 400\,km height to be 1600\,W\,m$^{-2}$ (550\,W\,m$^{-2}$ for $10<\nu<31\,$mHz, and 13\,W\,m$^{-2}\,$mHz$^{-1}$ for $\nu>31\,$mHz), which is significantly larger than the estimate of F\&C ($< 438\,$W\,m$^{-2}$ for waves between 5 and 50\,mHz).

We feel that it is justified to reconsider the significance of the energy flux of (non-M)HD waves, for four reasons: (1) In the past, the contribution of internal-gravity waves has been underappreciated. They turn out to carry enough energy to balance the radiative losses of the chromosphere. (2) With Hinode data we find an acoustic flux 3-5 times larger than F\&C. We believe that their estimate of the upper limit is underestimated. (3) The effects of the line formation MTF seem to have been underestimated in the past. (4) The propagation behaviour of high frequency acoustic waves in the ``real'' solar atmosphere appears to be more complicated than commonly assumed.
 
\acknowledgements Hinode is a Japanese mission developed and launched by ISAS/JAXA, with NAOJ as domestic partner and NASA and STFC (UK) as international partners. It is operated by these agencies in co-operation with ESA and NSC (Norway). SOHO is a project of international cooperation between ESA and NASA.

\end{document}